\documentstyle[12pt,aaspp4,rotate,psfig]{article}

\def\DF{{DF}}
\def\Mestel{Mestel}
\def\eg{{\it e.g.}}
\def\etal{{\it et al.}}
\def\etc{{\it etc.}}
\def\ie{{\it i.e.}}
\def\vcirc{{v_{\rm circ}}}
\def\betagammasmall{  1/\beta + \gamma/\beta - \gamma/2 }
\def\fsing{f_{\rm s}}
\def\Qsing{Q_{\rm s}}
\def\Qs{Q_{\rm s}}
\def\fcutout{f_{\rm c}}
\def\LNi{ L_z^{\nuh} }
\def\LNo{ L_z^{\muh} }
\def\Lc{ L_{\rm c} }
\def\Rc{ R_{\rm c} }
\def\rh{ r_{\rm h} }

\def\rhoh{ \rho_{\rm h} }
\def\psih{ \psi_{\rm h} }
\def\rhob{ \rho_{\rm b} }
\def\rhod{ \rho_{\rm d} }
\def\half{ \frac{1}{2}}
\def\GammaFunctsombtpb{ \frac %
        { \Gamma \left[ \half \left( 1-\beta \right)  \right]  %
         \Gamma \left[ \half \left( 2+\beta \right)  \right] } %
         { \Gamma \left[ \half \left( 1+\beta \right)  \right]  %
         \Gamma \left[ \half \left( 2-\beta \right) \right] } }

\def\Omegap{\Omega_{\rm p}}
\def\cosec{\rm cosec}
\def\sec{\rm sec}
\def\ffrac#1#2{{\textstyle\frac{#1}{#2}}}
\newcommand{\shalf}{{\textstyle\frac{1}{2}}}

\def\spose#1{\hbox to 0pt{#1\hss}}
\def\lta{\mathrel{\spose{\lower 3pt\hbox{$\sim$}}
    \raise 2.0pt\hbox{$<$}}}
\def\gta{\mathrel{\spose{\lower 3pt\hbox{$\sim$}}
    \raise 2.0pt\hbox{$>$}}}

\def\nuh{{\tilde \nu}}
\def\muh{{\tilde \mu}}
\def\betah{{\tilde \beta}}
\def\alphah{{\tilde \alpha}}

\slugcomment{Revised version to appear in \it The Astrophysical Journal}

\begin{document}

\title{The Stability of Disks in Cusped Potentials\footnote{Rutgers Astrophysics 
Preprint \#252}}
\author{J. A. Sellwood}
\affil{Department of Physics and Astronomy, Rutgers University, \\ 
136 Frelinghuysen Road, Piscataway NJ 08854 \\
sellwood@physics.rutgers.edu}

\author{N. W. Evans}
\affil{Theoretical Physics, Department of Physics, 1 Keble Road, Oxford, OX1 
3NP, UK \\
w.evans1@physics.oxford.ac.uk}

\begin{abstract}
We confirm that a high rate of shear at the center is able to stabilize an 
entire stellar disk against bar-forming ($m=2$) modes, irrespective of the dark 
halo density.  Our simulations of unstable power-law disks also yield the growth 
rate and pattern speed of the dominant lop-sided ($m=1$) mode in close agreement 
with that predicted by linear theory in our cleanest case.  The one-armed modes, 
which dominate models with extensive disks, can be controlled by tapering the 
outer disk.  Two-armed instabilities of hard-centered disks are more difficult 
to identify because they are easily overwhelmed by particle noise.  
Nevertheless, we have detected the predicted $m=2$ modes in simulations with 
very large numbers of particles.  Such bisymmetric instabilities in these disks 
are provoked only by sharp edges and are therefore easily eliminated.  

We have constructed a cool disk model with an almost flat rotation curve and a 
quasi-exponential density profile that is unambiguously stable.  The halo in 
this stable model has a large core radius, with the disk and bulge providing 
almost all the rotational support in the inner parts.

\end{abstract}

\section{Introduction}

The absence of strong bars in a significant fraction of disk galaxies has long 
been thought to present a serious problem for galactic dynamicists.  Many simple 
models of fully self-gravitating disks possess global, disruptive instabilities 
that form bars on a dynamical time scale (Miller, Prendergast \& Quirk 1970; 
Hohl 1971).  In an influential paper, Ostriker \& Peebles (1973) argued that 
unbarred galaxies were stable because they contained a large fraction of mass in 
a dynamically hot component which was unable to participate in collective 
instabilities.  Their paper has given rise to the impression that massive dark 
matter halos are required for the stability of all disk galaxies.

Toomre (1981) propounded a convincing mechanism for bar-forming modes.  Not only 
did he provide a physical understanding of the stabilizing effect of halos for 
disks with gently rising rotation curves, but he also argued that disks could be 
stabilized by making their centers inhospitable to density waves.  One way to 
achieve this is for the circular speed to remain high towards the center, which 
forces an inner Lindblad resonance (ILR) that cuts the feedback loop to the 
swing amplifier (\eg\ Binney \& Tremaine 1987 \S6.3).  Thus, even a fully 
self-gravitating disk can be stabilized against bisymmetric modes, much as 
Toomre's student Zang (1976) had found for the infinite, $V=\,$const.\ disk 
(here abbreviated as the Mestel disk, following the somewhat loose designation 
by Binney \& Tremaine).

Toomre's prediction was almost immediately challenged by Efstathiou, Lake \& 
Negroponte (1982), who found no evidence for stabilization by hard centers in an 
extensive set of numerical simulations.  The contradiction between their 
numerical results and Toomre's prediction was resolved by Sellwood (1989) who 
showed that large-amplitude disturbances in highly responsive disks could 
saturate the ILR, trapping particles in a large-scale bar similar to that which 
would have formed without the dense center.  Disturbances of sufficient 
amplitude to trigger a bar in this non-linear manner may well have arisen from 
particle noise in the rather grainy simulations by Efstathiou \etal, 
particularly when the disk is highly responsive.  As a result, the stabilizing 
effect of a dense center has been often discounted, giving rise to the flawed, 
but frequently recited, argument that since maximum disks are unstable, real 
galaxies cannot be maximum.  (A galaxy model is described as having a ``maximum 
disk'' when the dark matter contribution to the central attraction in the inner 
regions is negligible compared with that of the luminous disk and bulge.)

Sellwood (1985, 1999) and Sellwood \& Moore (1999) report $N$-body simulations 
of counter-examples which appeared robustly bar-stable.  These realistic galaxy 
models having cool, maximal disks with dense centers, are too complicated for 
their stability to be checked by linear analysis, however.  The simplest models 
with hard centers are disks in power-law potentials, and their linear stability 
properties have already been calculated.  For example, the global, 
small-amplitude modes in the \Mestel\ disk were already determined by Zang 
(1976) who found, rather surprisingly, that bisymmetric ($m=2$) modes were 
easily stabilized even when the disk was very cool.  Zang's result also holds in 
other disks with arbitrary power-law rotation curves (Evans \& Read 1998a,b).  
Unfortunately, all these power-law models suffer from strong one-armed 
instabilities instead, provoking Zang to conclude ``we may merely have jumped 
from the frying pan into the fire.''  Toomre (1981) later pointed out that $m=1$ 
instabilities can be controlled by reducing the disk surface density without 
changing the potential, much in the same manner as bars can be avoided in 
soft-centered disks by making them sub-maximal.

Our objectives in this paper are two-fold.  To confirm the extraordinary 
bisymmetric stability of the power-law disks, and to show that the lop-sided 
modes can also be quelled by tapering the outer disk.  Such tapering again 
reduces the surface density everywhere, but changing the shape of the radial 
surface density profile from a power law to roughly exponential increases the 
peak disk contribution to the central attraction, preserving a maximum disk.

We begin by demonstrating reasonable agreement between results from $N$-body 
simulations and predictions from the linear stability analyses of power-law 
disks.  Previous work has shown that the eigenmodes predicted by linear theory 
can be successfully reproduced in $N$-body experiments.  These tests were for 
the Kalnajs disks (Sellwood 1983), the Kuz'min-Toomre disk (Sellwood \& 
Athanassoula 1986), and the isochrone (Earn \& Sellwood 1995).  However, all 
three tests were of models with uniform density cores and therefore finite 
central frequencies, and we quickly discovered that simulations of power law 
disks require much more care in order to reproduce the predicted modes.  While 
we are ultimately able to claim success, our difficulties arise directly from 
this physical difference and illustrate interesting aspects of disk dynamics.

Bisymmetric instabilities in power-law disks are provoked by a sharp inner 
cut-out.  Not only are we able confirm these instabilities, but we also verify 
that an almost full-mass disk with a sufficiently gentle inner edge is 
bisymmetrically stable.

We demonstrate that the lop-sided instability can be removed by tapering away 
the massive, extended outer disk.  We have been able to adapt a power-law disk 
to construct a completely stable and reasonably realistic galaxy model in which 
the disk provides most of the central attraction in the inner parts.  This model 
has a quasi-exponential disk and an almost flat rotation curve that arises from 
a combination of the massive disk, a central bulge, together with a halo having 
very low central density.  We show, using both linear stability theory and 
numerical simulations, that it has no global instabilities whatsoever.

\section{The Equilibria}
\label{sec:equilibria}

Motivated by the near-flatness of galactic rotation curves, we here study a 
family of models which possess rotation speeds that vary as a power of the 
radius.  We break their scale-freeness by altering their surface density 
profiles using inner and outer tapers without changing the gravitational field.

The circular velocity is
\begin{equation} 
         \vcirc^2 = \left( R_0 \over R \right) ^\beta v_0^2,
        \label{eq:circvel}
\end{equation} 
where $v_0$ is the velocity at the reference radius $R_0$.  When $\beta =0$, the 
rotation curve is completely flat; otherwise, the rotation curve is rising if 
$\beta <0$ and falling if $\beta >0$.  The gravitational potential in the plane 
of the disk is
\begin{equation} 
     \psi (R) =   \cases{ \displaystyle \frac{v_0^2}{\beta} \left( R_0 \over 
R \right)^\beta & $\beta \neq 0$ \cr
     \displaystyle -v_0^2 \ln \left( {R \over R_0}\right) & $\beta = 0.$ \cr}
\label{eq:potential}
\end{equation} 
The surface density of a full-mass disk is
\begin{equation} 
         \Sigma(R) = {v_0^2 \over 2\pi GR_0{\cal L}(\beta)}
            \left( {R_0 \over R} \right)^{1+\beta},
        \label{eq:surfden}
\end{equation}
with
\begin{equation}
     {\cal L}(\beta) =         \GammaFunctsombtpb. 
\end{equation}
Note that ${\cal L}(\beta) = 1$ for the \Mestel\ disk, since $\beta = 0$.

We will need the swing-amplification parameter $X$, defined by Toomre (1981) for 
an $m$-armed disturbance as
\begin{equation} 
         X = {2 \pi R \over m \lambda_{\rm crit}} 
           = {(2-\beta){\cal L}(\beta) \over m}.
        \label{eq:Xdef}
\end{equation}
Here $\lambda_{\rm crit}$ is given by 
\begin{equation} 
        \lambda_{\rm crit} = {4 \pi^2 G \Sigma \over \kappa^2},
        \label{eq:lcrit}
\end{equation} 
which local theory predicts is the longest unstable wavelength of axisymmetric 
Jeans instabilities in a purely rotationally supported disk (Toomre 1964).  
Here, $\kappa$ is the radial epicyclic frequency, as usual.

The distribution functions (hereafter \DF s) depend on both isolating integrals 
of motion: the energy per unit mass $E$ and the specific angular momentum $L_z$. 
 The \DF s are
\begin{equation}
     \fsing(E,L_z) =  \cases 
         {\tilde{C}  L_z^{\gamma} | E |^{\betagammasmall} & $\beta \neq 0$ \cr  
          \tilde{C} L_z^\gamma \exp \left[ -(\gamma+1){E /v_0^2} \right], & 
$\beta = 0$ \cr}
        \label{eq:fsing}
\end{equation}
as given by Evans (1994) for the general case and by Toomre (1977, see also 
Binney \& Tremaine 1987) for the \Mestel\ disk.  In both these formulae, 
$\tilde{C}$ is a normalization constant, whose value can be found in Evans 
(1994) or Evans \& Read (1998a).  The velocity anisotropy constant $\gamma$ 
prescribes the radial velocity dispersion $\sigma_u$ of the stars:
\begin{equation}
\sigma_u^2 = {{v_0^2} \over {1+\gamma+2\beta} } \left( {R_0 \over R }
\right) ^{\beta}.
\end{equation}
For these full-mass, self-similar disks, the stability parameter
\begin{equation}
\Qsing = \frac {2 \pi {\cal L}(\beta)}{3.36} 
               \sqrt{\frac{2-\beta}{1+\gamma+2\beta}}
                                            \label{eq:localsigmin},
\end{equation}
is independent of radius.  Local theory (Toomre 1964) tells us that the disk is 
everywhere stable to axisymmetric modes provided $\Qs \ge 1$.

The \DF s~(\ref{eq:fsing}) generate the mass density implied by the 
gravitational potential through Poisson's equation.  A self-similar disk with no 
preferred scale has unusual stability properties (Goodman \& Evans 1999), and in 
any case differs greatly from the quasi-exponential profiles of real galactic 
disks.  We therefore taper the disks with both inner and outer cut-outs, very 
much akin to Zang's (1976) pioneering investigation.  The cut-out mass is still 
present, in the sense that it contributes to the forces experienced by the 
remaining stars, but it is not free to participate in any perturbations.  The 
actual \DF\ we use is that for the self-similar case multiplied by a taper 
function $H(L_z)$:
\begin{equation}
        \fcutout (E,L_z) = H(L_z) \fsing(E,L_z),
        \label{eq:fcutouteq}
\end{equation}
where
\begin{equation}
	H(L_z) = { \LNi \over 
 \left[\LNi + \left( v_0 R_0 \right)^{\nuh}\right]}
 { \Lc^{\muh} \over \left[ \LNo + \Lc^{\muh} \right]}. 
        \label{eq:hfactors}
\end{equation}
The index $\nuh$ controls the inner cut-out, while $\muh$ determines the outer 
taper; the larger the value of either index, the more abrupt the tapering.  The 
outer taper is centered on a characteristic angular momentum $\Lc$, or radius 
$\Rc$, where $\Lc = \Rc \vcirc (\Rc)$.  In simulations, we are free to choose 
whatever $\nuh$ and $\muh$ we wish. However, the normal mode analysis is 
available only for the special values
\begin{equation}
\nuh = \nu {2 + \beta\over 2- \beta}, \qquad \muh = \mu {2 + \beta
\over 2 - \beta},
\end{equation}
where $\nu$ and $\mu$ are integers.  These preferred values greatly simplify the 
contour integral in the normal mode analysis (see Appendix C of Evans \& Read 
1998a).  Note that for the important $\beta =0$ case, there is no difference 
between $\nuh$ and $\nu$, $\muh$ and $\mu$.

The equilibrium and linear stability properties of these models are discussed in 
more detail in Evans \& Read (1998a,b).  For the remainder of this paper, we 
adopt a system of units in which $G = v_0 = R_0 = 1$.  Unless otherwise stated, 
the models in our simulations have $\nu=4$, $\mu=6$, $\Rc=10$ and $\Qs=1$.

\section{Numerical Techniques}

\label{sec:numerics}

\subsection{Aliasing}

The acceptable agreement reported in \S\S \ref{sec:resultstwo} \& 
\ref{sec:resultsone} between linear theory predictions and our results was 
reached only after many failures.  In the light of the successful experience of 
Earn \& Sellwood (1996), we began by using the smooth-field-particle (SFP) 
method to determine forces.  While less efficient than particle-mesh (PM) 
methods, particularly for larger numbers of basis functions, it does not require 
forces to be softened at all.

Results with first 10 -- 25 members of the $k_{\rm AJ}=7$ set of Abel-Jacobi 
functions (Kalnajs 1976) were appalling.  The $m=2$ coefficients grew in a 
largely incoherent manner at a rate roughly ten times higher than that 
predicted, with ``pattern speeds'' which seemed quasi-continuous over a broad 
swathe of values.  Non-axisymmetric features grew somewhat less rapidly in 
simulations with larger numbers of functions, but still no single ``mode'' stood 
out.

After much experimentation with different time steps, particle numbers, \etc, 
which yielded little in the way of improved behavior, we switched to Bessel 
functions as our basis set.  Unlike the previous discrete basis, a Hankel 
transform has a continuous distribution of radial wavenumbers, $k$, which in 
practice has to be represented by a discrete set of values.  While our first 
experiments revealed no significant improvement, we soon noticed that the 
average rate of growth correlated with the value of $\delta k$ adopted; \ie\ the 
smaller $\delta k$, the lower the growth rate we observed.  Furthermore, 
inspection of the Fourier transform of the best fitting ``mode'' revealed a 
leading signal in excess of that predicted by linear theory, which gradually 
diminished as $\delta k$ was reduced.

This behavior was a strong hint of aliasing.  The transforms of somewhat tightly 
wrapped waves, both trailing and leading, are inadequately represented by a 
loose spacing in $\delta k$.  Thus, as noise induced waves sheared towards the 
trailing side, the gravitational field acquired components of enhanced {\it 
leading\/} waves.  This purely numerical feedback from trailing to leading was 
clearly responsible for the unphysically rapid growth of non-axisymmetric 
structure in our models.  The required tighter spacing in $k$ led to rapidly 
escalating cpu time requirements.  Computing significantly more than 100 
functions, as our results indicated would be needed, rendered this method 
uneconomic.  (It may perhaps be possible to reproduce the predicted behavior 
with a much smaller number of functions from some other basis set.)

Reverting to a PM code immediately avoided such problems, and led to much more 
physically reasonable behavior.  However, particle number, grid size, time step 
and softening all needed to be substantially improved over values that seemed 
adequate in previous work before we were able to obtain results in reasonable 
agreement with the predictions.

\subsection{Adopted method}

We used a 2-D polar grid (Sellwood 1981) to determine the forces from the 
particles.  This PM method is highly efficient but does require forces to be 
softened, if only slightly, in order to prevent relaxation (Rybicki 1972).

We are here concerned mostly with linear instabilities, for which the radial 
distribution of mass is assumed to remain unchanged.  For these cases, 
therefore, we need not evaluate the axisymmetric part of the field from the 
particles, and use instead the analytic central attraction of the infinite 
self-similar model.  In fact, as the different sectoral harmonics of the 
perturbing forces from small amplitude disturbances are also assumed to be 
decoupled in linear theory, the particles in our linearized simulations 
contribute to just a single azimuthal Fourier component of the force field.

As these power-law models have a large range of orbital frequencies from the 
center to the edge, it is advantageous to employ a range of time steps.  The 
scheme adopted (Sellwood 1985) divides the model into a number (here 5) of 
radial zones, with the time step increasing by a factor 2 between zones.  Forces 
arising from all the particles in one zone are evaluated at every step and 
stored separately.  They are combined to find the total force from the entire 
disk, using linear interpolation between the nearest pair of times whenever 
forces are needed in the inner zones at a time not coincident with an outer 
step.  This scheme is exactly time-reversible, therefore.

As even this short step becomes inadequate near the center, we further subdivide 
the time step whenever a particle moves closer to the center than a ``guard 
radius.''  For such a particle, we evaluate the force acting, which is anyway 
dominated by the fixed axisymmetric component this close to the center, by 
assuming that no others move until the next step of the innermost zone.  The 
guard radius is always equal to $1/6$, within which the time step is shorter by 
a factor 10.

The principal result from each simulation is an estimate of the eigenfrequency 
of the dominant mode.  We determine these frequencies using the mode fitting 
procedure described in detail by Sellwood \& Athanassoula (1986); see also Earn 
\& Sellwood (1995).  As usual, our estimated uncertainties indicate the entire 
range that encompasses all credible fits to subsets of the data from that run, 
with several scaling factors.

\subsection{Selection of Particles}
\label{sec:choosingparticles}

The implementation of quiet start procedures for galaxy simulations has 
developed from the initial ideas outlined by Sellwood (1983), and the latest 
prescription is described in appendix B of Debattista \& Sellwood (2000).  Since 
the \DF\ is expressed as a function of the two classical integrals, $f(E,L_z)$, 
we aim to create a sample of particles which has a density in this space as 
close as possible to $f$.  We proceed by slicing accessible $(E,L_z)$ space into 
$j$ small areas in such a way that the integral of the \DF\ over each area 
encloses a fraction $1/j$ of the total active mass.  We then select (not 
entirely at random) a point within each of these areas to determine the 
$(E,L_z)$ values for an orbit.  For each orbit selected in this way, we then 
choose $l$ radial phases to determine both the initial radial position and the 
azimuthal and radial velocity components of each particle.  We finally place $n$ 
particles having these three phase space coordinates at equally spaced azimuthal 
locations.  Thus, the total number of particles $N=jln$.  While as deterministic 
as we could make it, this procedure requires us to chose the shapes of each of 
the $j$ areas in $(E,L_z)$-space, and the values of $l$ and $n$.  As a result, 
we are able to create many different samples from the \DF\ by varying these 
parameters.

It seemed possible that the behavior of the modes might depend sensitively on 
the representation of the inner cut-out (see \S \ref{sec:discussiontwo}).  If 
this were true and all particles have equal masses, discreteness effects would 
ultimately become important in this region no matter how large the number of 
particles employed.  We therefore experimented with packing more particles into 
this region while retaining the desired density distribution by assigning 
individual masses.  In most cases, we selected particles as if there were no 
inner cut-out, but we also tried to bias the \DF\ with an extra factor (a 
function either of $E$ or of $L_z$) in order to place a still greater fraction 
of the particles in the inner disk.  While seemingly desirable, this strategy 
actually back-fired, especially for models with declining rotation curves.  By 
packing many low-mass particles in the central region, we have fewer left to 
represent the massive outer part of the disk.  Results for declining rotation 
curves were clearly better when all particles had equal mass, hinting that noise 
in the massive part of the disk, rather than representation of the inner edge, 
is the principal source of error.

\section{Bisymmetric Modes}

\subsection{Comparison with Linear Stability Theory}
\label{sec:resultstwo}

Zang (1976) showed that the full-mass \Mestel\ disk is stable to $m=2$ modes 
unless the central cut-out index $\nu>2$.  For a model with $\nu=4$ and 
$\Qsing=1$, he predicted the eigenfrequency of the dominant $m=2$ mode to be 
$m\Omegap+i\,s = 0.879+0.127i$; where $s$ is the growth rate and $\Omegap$ the 
pattern speed.  As simulations of an infinite disk are impossible, we add an 
outer taper with index $\mu = 6$ centered at $\Rc=10$, which changes the 
predicted frequency very slightly to $2\Omega_p+i\,s = 0.879+0.128i$.  In order 
to keep the radial extent finite, we impose an additional energy truncation such 
that no particle's orbit takes it outside $R=15$.  This extra truncation removes 
little mass and is therefore unlikely to affect the eigenfrequency further.

\begin{table}[t]
\begin{center}
\begin{tabular}{|ll|}
\tableline 
\rule{0pt}{12pt}
Parameter & Value \\ 
\tableline
Number of particles ($N$) & $2.5 \times 10^6$ \\
Grid size & $200 \times 256$ \\
\qquad $\Delta\phi$ & $2\pi/256$ \\
\qquad $\Delta R/R$ & $\simeq 0.0248$ \\
Softening length ($\epsilon$) & $R_0/60$ \\
Time step (inner zone) & $0.005R_0/V_0$ \\
\tableline
\end{tabular}
\caption{Standard numerical parameters in our particle-mesh simulations.}
\label{table:jerrysparameters}
\end{center}
\end{table}

\begin{figure}[p]
\centerline{\psfig{file=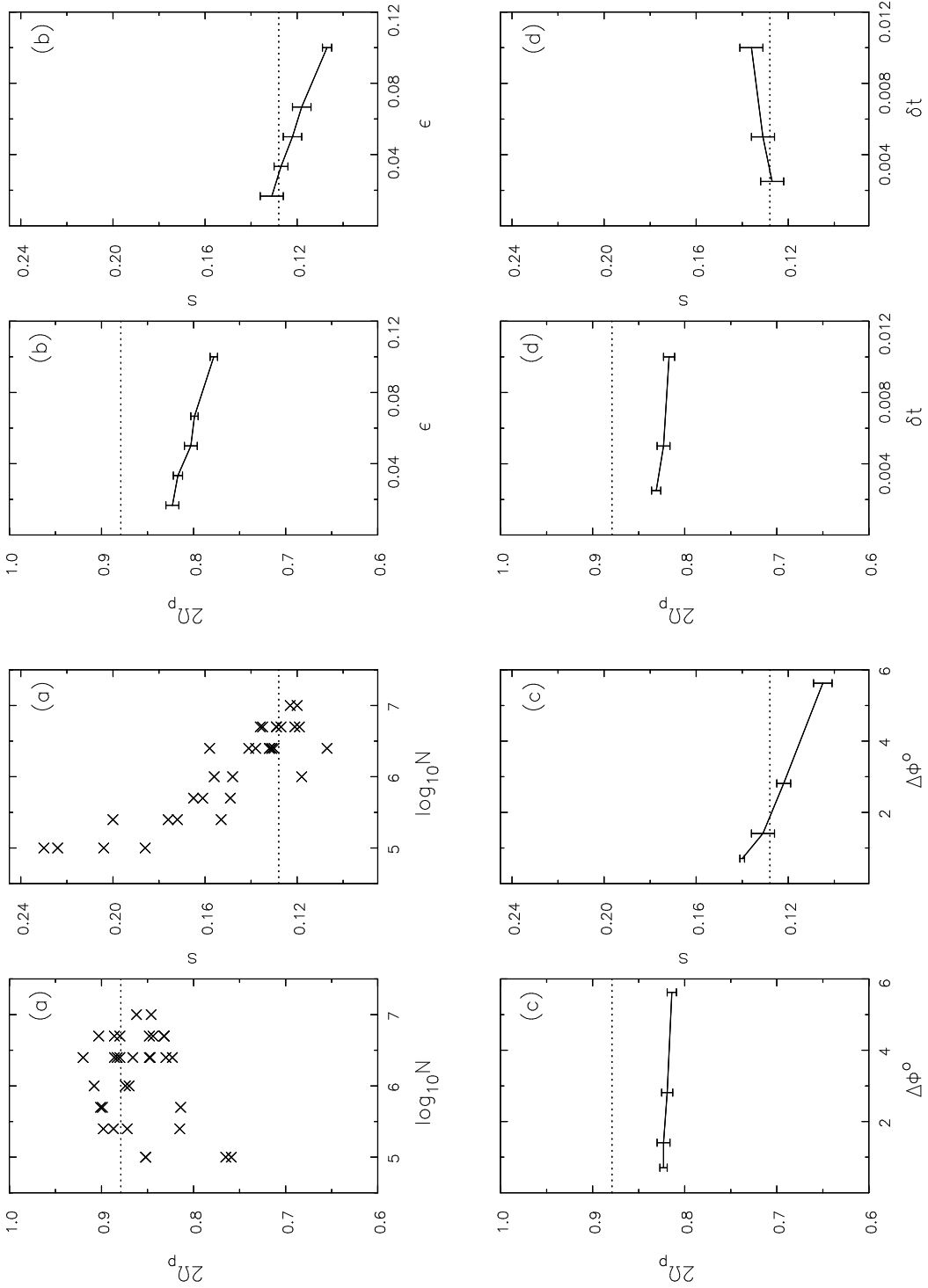,width=0.9\hsize,angle=0}}
\caption{The effects of varying the numerical parameters on the pattern speed 
and growth rate of the dominant bisymmetric mode of the \Mestel\ disk with 
$\nu=4$, $\mu=6$ and $\Rc = 10$.  Our standard values are given in 
Table~\ref{table:jerrysparameters}.  The parameters that are varied are: (a) 
number of particles $N$ (no error bars shown), (b) softening length $\epsilon$, 
(c) grid resolution $\Delta \phi$ and (d) time step $\delta t$. The prediction 
from linear theory is shown as a dotted line.}
\label{fig:jerrysmodes}
\end{figure}

Figure~\ref{fig:jerrysmodes} summarizes the results from many simulations of 
this one model as the numerical parameters are varied.  Each pair of panels 
shows how the measured pattern speed and growth rate vary as one numerical 
parameter is changed from the ``standard'' values given in 
Table~\ref{table:jerrysparameters}.  The horizontal dotted lines indicate the 
frequency predicted from linear theory.

As the number of particles is increased (panels a), the measured frequencies 
clearly approach the predicted value near $N=2.5 \times 10^6$.  The scatter 
results from making many different selections of particles from the \DF\ for 
each value of $N$.  The steady trends seen in panels (b), (c) \& (d), on the 
other hand, result from models which all began with exactly the same sample of 
particles.

The data we fit to estimate the eigenfrequency are never well approximated by a 
single mode, and all our fits require a second ``mode.''  Unlike the dominant 
mode, however, the second wave does not have a coherent frequency throughout its 
growth, neither is its frequency similar from runs with different samples from 
the DF.  Since amplified particle noise can be reasonably well approximated by a 
coherent wave for short time periods, our fitting procedure is never able to 
separate noise completely from the true mode.  The scatter in panels (a) and the 
lack of scatter in the other panels strongly suggest that our measurements are 
most affected by the noise spectrum in the simulation, which is changed with 
every new sample drawn from the \DF.  We discuss particle noise in more detail 
in \S\ref{sec:noise}.

The trends all show that the results have yet to ``converge'' as each parameter 
is refined.  Our standard parameters (Table~\ref{table:jerrysparameters}) are 
not perfect, but seem adequate.  It may seem worrisome that the trends in panels 
(b), (c) \& (d) are not converging to the prediction.  We stress, however, that 
these results were obtained from a single sample of particles, and the true 
error bars on these results should be enlarged considerably to take account of 
the $\sim 5-10$\% scatter for this $N$ in panels (a).

\begin{figure}[t]
\centerline{\psfig{file=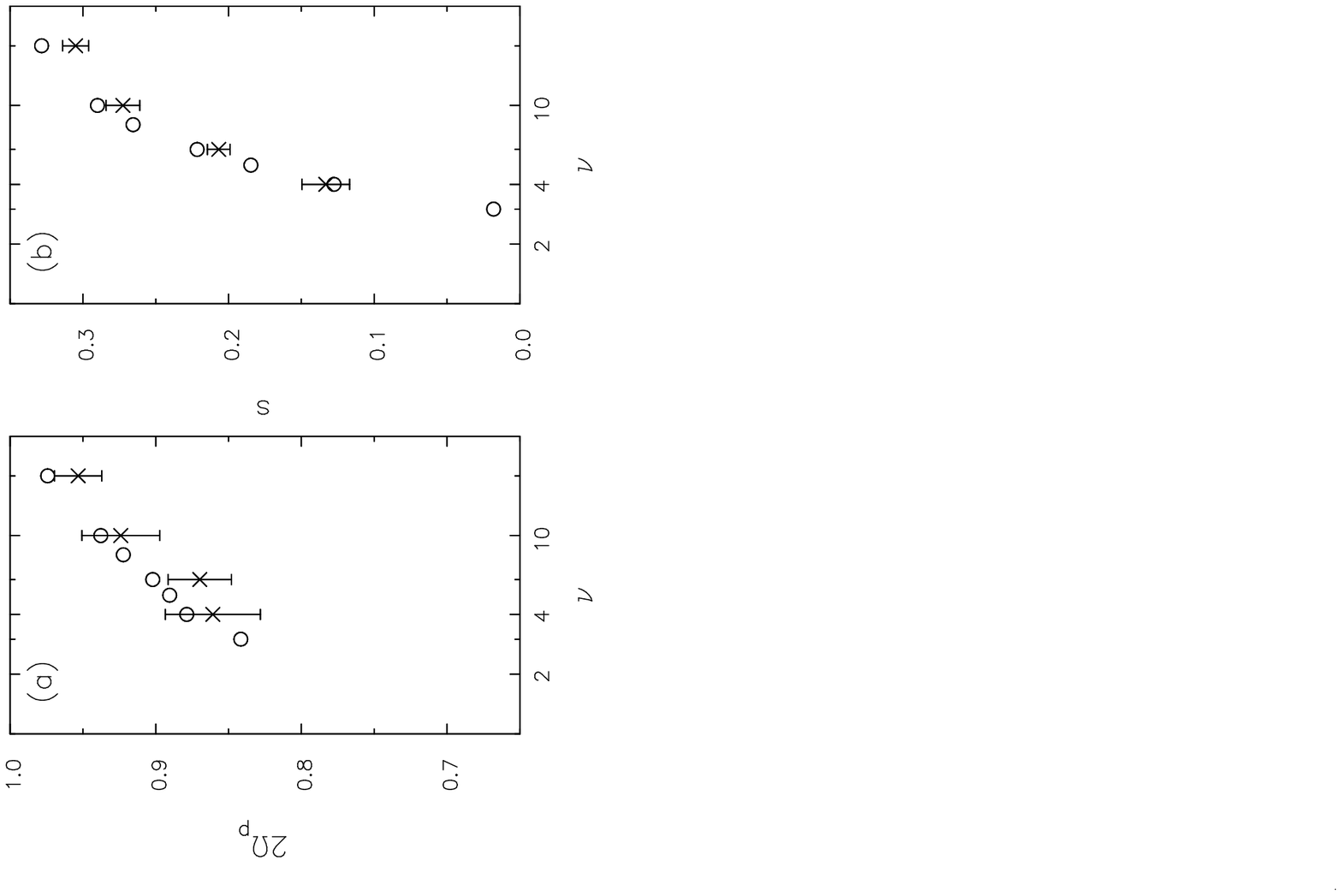,width=0.7\hsize,angle=270}}
\caption{The effects of varying the inner cut-out index $\nu$ on the pattern 
speed and growth rate of the dominant bisymmetric mode of the \Mestel\ disk.  
The circles show the linear theory predictions and the error bars show the 
dispersion from several simulations with different samples drawn from the DF for 
each case.}
\label{fig:sharp}
\end{figure}

The scatter in panels (a) seems to decrease as $N$ rises suggesting that the 
predicted values might be reliably reproduced in still larger calculations, as 
they should.  We obviously should expect our higher quality results to cluster 
around the prediction, which we claim they do.

The predictions and results for several values of the inner cut-out index $\nu$ 
are shown in Figure~\ref{fig:sharp}.  The error bars on the data points show the 
1-$\sigma$ dispersion (not the standard error) from a number ($\geq4$) samples 
from the DF in each case.  All these simulations used the standard numerical 
parameters given in Table~\ref{table:jerrysparameters}.  We find reasonable 
agreement between linear theory and the simulation results, but generally far 
below the level Earn \& Sellwood (1995), who employed one twentieth the number 
of particles, were able to achieve for the isochrone disk.  Even with the larger 
$N$ used here, no bisymmetric mode, however strongly growing, stands out clearly 
from the noise in the manner observed routinely in simulations of soft-centered 
disks.  The values we obtain have a tendency to underestimate the prediction, of 
$\Omega_p$ in particular, which we find bothersome; we have been unable to 
eliminate this small systematic difference and believe residual noise to be the 
most likely culprit.

\begin{table}[t]
\begin{center}
\begin{tabular}{|c|c|cc|}
\tableline 
\rule{0pt}{12pt}
& Predicted Frequency & Observed & Observed \\
$\beta$ & $2\Omega_p + i\,s$ & Real part & Imaginary part 
\\ 
\tableline
 -0.25 & $0.855 + 0.191i$ & $0.840\pm 0.018$ & $0.171\pm 0.020$ \\ 
       &                  & $0.831\pm 0.001$ & $0.186\pm 0.003$ \\ 
       &                  & $0.832\pm 0.004$ & $0.189\pm 0.010$ \\ 
       &                  & $0.864\pm 0.001$ & $0.198\pm 0.005$ \\ 
\tableline
0.25 &  $0.920 + 0.060i$ & $0.757\pm 0.011$ & $0.097\pm 0.017$ \\ 
     &                   & $0.865\pm 0.038$ & $0.085\pm 0.011$ \\ 
     &                   & $0.805\pm 0.013$ & $0.102\pm 0.010$ \\ 
     &                   & $0.865\pm 0.002$ & $0.086\pm 0.004$ \\ 
\tableline
\end{tabular}
\end{center}
\caption{Predicted and estimated eigenfrequencies of $m=2$ modes in power-law 
disks.  The upper panel refers to the models with rising rotation curves ($\beta 
= -0.25$), the lower panel to falling rotation curves ($\beta = 0.25$).}
\label{table:resultstwo}
\end{table}

Our results for two different power-law models, one with rising and the other 
with falling rotation curves, are summarized in Table~\ref{table:resultstwo}, 
together with the predictions from linear theory, for $\nu=4$ in each case.  The 
agreement for the model with the rising rotation curve ($\beta = -0.25$) is 
satisfactory, but that for the falling rotation curve model is much poorer.  
Further investigation seems to indicate that the number of particles needed for 
good agreement rises with increasing $\beta$.  One part of the reason for this 
behavior was traced to the folly of filling the inner cut-out with a large 
fraction of low-mass particles, as noted in \S \ref{sec:choosingparticles}.  But 
this is clearly not the full story, since the results given in the Table 
\ref{table:resultstwo} (for this case only) are for equal mass particles.  While 
slowly-growing modes are clearly more difficult to reproduce, \S\ref{sec:noise} 
gives a further reason why the falling rotation curve case is worse.

\subsection{Mode mechanism}
\label{sec:discussiontwo} 

The reason the modes of soft-centered disks are so much easier to reproduce than 
those of the power-law disks stems from the difference in reflection mechanisms 
between the two types of disk.  In both cases, they are swing-amplified inner 
cavity modes (Toomre 1981) in which in-going trailing waves return as out-going 
leading waves providing positive feedback to the amplifier.  A wave-train in a 
soft-centered disk, such as the isochrone, reflects off the disk center and 
returns as a leading wave without attenuation.  The centers of the power-law 
disks, on the other hand, cannot support density waves for two reasons: first, 
frequencies become arbitrarily large as $R \rightarrow 0$, which ensures an ILR 
at some radius for all pattern speeds, and second, the inner cut-out causes $Q$ 
to rise steeply towards the center.  Feedback in this disk is through reflection 
off the inner cut-out which probably returns some signal in both the 
short-leading and long-trailing wave channels.  It is also possible that some 
fraction of the in-coming wave is absorbed by the imperfectly shielded ILR, 
especially when the resonance is broadened by the rapid growth of the mode.  The 
sharper the inner cut-out, the stronger the reflected leading wave becomes and 
the more vigorous the mode.

Thus, unlike for the soft-centered disks, only a fraction of the in-going 
trailing wave returns as out-going leading wave to provide feedback to the swing 
amplifier.  The fraction returning as long-trailing waves also shears to 
short-trailing near corotation (\eg\ Mark 1976), but with mild amplification 
only when $Q \lta 1.2$.  The effect of this ``leakage'' from the main 
swing-amplified feedback loop is to slow the rate of growth of the mode.  The 
response of the disk to noise, on the other hand, is unaffected by feedback and 
is therefore proportionately more troublesome in the power-law disks.

\subsection{Amplified particle noise}
\label{sec:noise}

When the reflected leading signal is weak, as in disks with gentle cut-outs, the 
predicted mode is easily submerged by particle noise, but particle noise remains 
an unusually severe problem even in disks with sharp cut-outs.  The leading 
component of the noise spectrum produces apparently growing non-axisymmetric 
features that obscure our view of the underlying true instability.

\begin{figure}[t]
\centerline{\psfig{file=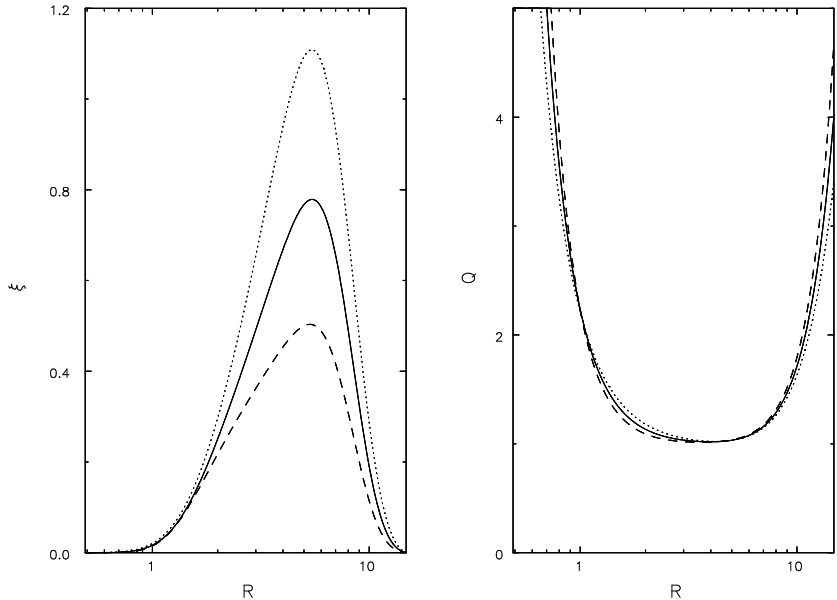,width=.7\hsize,angle=270}}
\caption{Left: Radial variation of the fraction of mass per square $\lambda_{\rm 
crit}$, $\xi$, for the three different power-laws: the dashed curve is for 
$\beta = 0.25$, unbroken for $\beta = 0$ and dotted for $\beta = -0.25$.  The 
number of particles per $\lambda_{\rm crit}^2$ is $N$ times the quantity 
plotted.  Right: The radial variation of $Q$ for these three disk models.}
\label{fig:grainy}
\end{figure}

\begin{figure}[p]
\centerline{\psfig{file=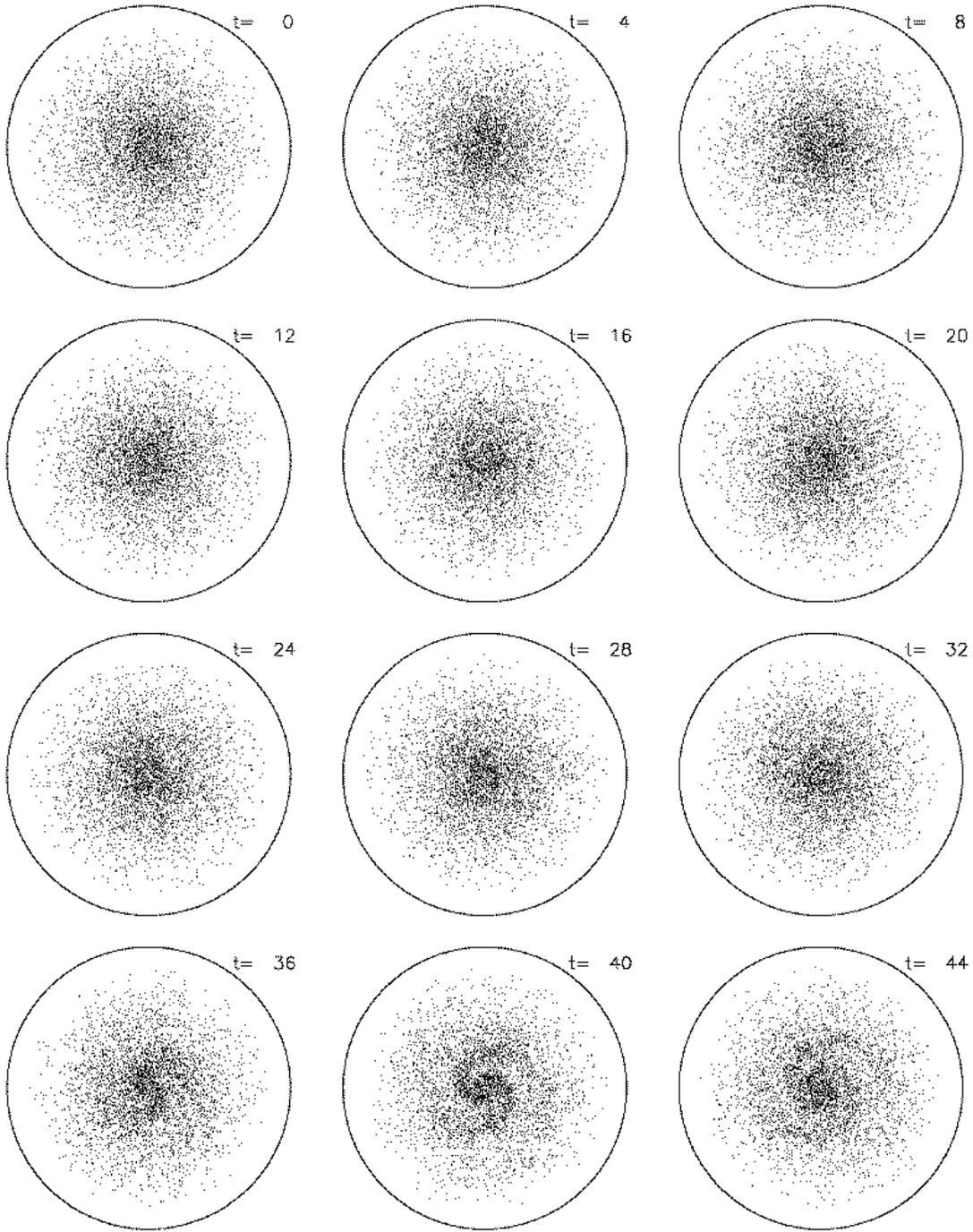,width=.9\hsize}}
\caption{Snapshots showing the position of one particle in 500 from a noisy 
start simulation with $N=2.5$ million particles in which perturbation forces 
were restricted to $m\leq4$ and $m\neq1$.  The radius of the circles is 
$17.2R_0$ and the times are in units of $R_0/v_0$.  The model has no true global 
instabilities.  Swing-amplified noise quickly produces large-amplitude patterns 
and later a bar even for this large $N$.}
\label{fig:noisy}
\end{figure}

When particles are distributed randomly, the level of particle noise depends 
both on the number of particles per square $\lambda_{\rm crit}$ (\eg\ Toomre \& 
Kalnajs 1991) and the vigor of the swing amplifier.  The maximum growth that the 
swing amplifier can produce depends on several properties of the disk: the 
parameters $Q$ and $X$, as well as the shear rate $\Gamma = 1 + \beta/2$.  We 
find that the expected amplification remains roughly constant as $\beta$ varies. 
 (We held $\Qs$ constant and the effects of changing both $\Gamma$ and $X$ 
largely cancel.)

The graininess of the disk, on the other hand, does depend on the power-law 
index, as shown in Figure \ref{fig:grainy}.  When all particles have equal mass, 
the number of particles per square $\lambda_{\rm crit}$ is $N\xi$ where $\xi$ is 
the ratio of mass per $\lambda_{\rm crit}^2$ to the total active mass in the 
simulation.  The regions of the disk where $\xi$ is small have high $Q$ (as 
shown in the right-hand panel) and do not therefore cause difficulties.  Note 
that the peaks of these curves are of order unity, so we have millions of 
particles per $\lambda_{\rm crit}^2$ in our largest calculations.

A quiet start is designed to reduce the seed amplitude of non-axisymmetric 
features in order to be able to observe linear growth over many $e$-folds.  By 
placing a number of particles at equal angular spacing on rings and restricting 
forces to low order sectoral harmonics, initial disturbance forces are many 
orders of magnitude below those that would have arisen from random initial 
positions.  Such an arrangement is not entirely noise free, however.  We provide 
an initial seed noise spectrum by nudging every particle away from perfect 
azimuthal spacing by a randomly chosen fraction of $0.002$ radians, or about 
$0.1^\circ$.  This initial noise seeds the growth both of the desired global 
mode (if one is present) as well as a swing-amplified response.  The resulting 
enhanced disturbance forces further distort the regular particle arrangement, 
creating enhanced density fluctuations which are further amplified, and so on.  
Thus the quiet start is itself unstable, and non-axisymmetric amplitudes rise 
until the distribution of particles is essentially random.  The rate at which 
swing-amplification drives the break up again depends on the responsiveness of 
the disk as well as $N\xi$.

In order to quantify the rate at which quiet starts disrupt, we have run 
simulations of our usual \Mestel\ disk but with the inner cut-out index $\nu=2$. 
 Zang (1976) concluded that the \Mestel\ disk is well stable to modes with 
$m\neq1$ in this case and our addition of an outer taper does not alter his 
finding.  Swing-amplification is still present in this ``stable'' case, however. 
 The amplitude of $m=2$ disturbances in a quiet start simulation grew in a 
quasi-exponential manner at an approximate rate of 0.06, but we could find no 
evidence for coherent modes.  This simulation confirms that swing-amplification 
from the initial noise causes a rapid break-up of the quiet start even when the 
equivalent smooth disk has no instability.

Amplified particle noise is the reason that growth rates appear to increase as 
$N$ decreases in Figure~\ref{fig:jerrysmodes}(a).  Only when $N \gta 10^6$ is 
the break-up rate slow enough to have little effect on the growth rate of the 
underlying mildly growing normal mode.  For fixed $N$, the $\beta = 0.25$ case 
has the smallest number of particles per $\lambda_{\rm crit}^2$ (dashed curve in 
Figure \ref{fig:grainy}), and amplified particle noise should be the greatest 
nuisance for this case, as we have found.  It should be noted that quiet starts 
neither create nor alleviate this difficulty because they reduce the amplitudes 
of {\it both\/} the mode and the noise in proportion.

\subsection{Confirmation of bisymmetric stability}
The severity of the noise problem is illustrated in Figure \ref{fig:noisy} which 
shows snapshots from a noisy start simulation of the linearly stable model; 
numerical parameters were as given in Table \ref{table:jerrysparameters} except 
that we placed each of the 2.5 million particles at a randomly chosen azimuth 
and included forces from the density sectoral harmonics 0 through 4, with only 
$m=1$ omitted.  Despite there being no true global instability, after just 40 
dynamical times (or one orbital period at $R \sim 6$), swing amplified particle 
noise has already produced a visible spiral pattern of sufficient amplitude to 
alter the density distribution and heat the inner disk.

Noise-driven activity in this lively $Q_s=1$ disk is so strong, even for this 
large number of particles, that the model later develops a strong bar.  The bar 
does not result from a linear instability, but through saturation of the ILR, or 
non-linear trapping, as discussed by Sellwood (1989).  The swing-amplifier is 
tamed by raising $\Qs$ and a further long simulation has confirmed the global 
stability of a similar model with $Q_s=1.5$.

\section{One-armed modes}
\label{sec:resultsone}

\subsection{Comparison with Linear Stability Theory}

Linear theory (Zang 1976; Evans \& Read 1998b) predicts a dominant pair of $m=1$ 
modes with closely spaced frequencies for cool power-law disks.  For our 
centrally cut-out and outer truncated $\Qs=1$ disks, the predicted frequencies 
for the dominant mode pair are given in Table~\ref{table:resultsone}.  These 
frequency differences are typical.  The Table also presents results from a 
number of separate simulations using different particle samples, but were 
otherwise identical for each disk and used the numerical parameters given in 
Table~\ref{table:jerrysparameters}.

In all our simulations of the \Mestel\ and power-law disks, the $m=1$ data were 
consistent with just a single growing mode having the indicated eigenfrequency, 
but we were able to find hints of the two predicted modes in one case.  
Separation of two such closely-spaced modes would require a long period of 
integration in the simulations, which cannot be continued indefinitely as the 
modes saturate after approximately 8 $e$-folds, or roughly 100 dynamical times.  
In the linear regime, therefore, they will have rotated relative to one another 
by only some 1.6 radians.  It seems reasonable that we should find growth rates 
consistent with that of the more rapidly growing mode, but it is curious that we 
sometimes find a pattern speed closer to that of the more slowly growing mode.  
The variations within a set of experiments using the same physical model 
probably arise from different relative amplitudes of the two modes seeded by the 
different initial particle noise.

\begin{table}[t]
\begin{center}
\begin{tabular}{|c|cc|cc|}
\tableline 
\rule{0pt}{12pt}
& \multicolumn{2}{c|}{Predicted frequencies} & Observed & Observed \\
$\beta$ & Mode 1 & Mode 2 & Real part & Imaginary part \\ 
\tableline
 -0.25 & $0.176 + 0.130i$ & $0.140 + 0.089i$
                  & $0.163 \pm 0.004$ & $0.128 \pm 0.004$ \\ 
           &   &  & $0.162 \pm 0.003$ & $0.125 \pm 0.002$ \\ 
\tableline
 0 & $0.189 + 0.118i$ & $0.173 + 0.089i$
                  & $0.172 \pm 0.002$ & $0.105 \pm 0.002$ \\ 
           &   &  & $0.169 \pm 0.003$ & $0.111 \pm 0.007$ \\
           &   &  & $0.173 \pm 0.005$ & $0.118 \pm 0.006$ \\
\tableline
0.25 & $0.213 + 0.100i$ & $0.198 + 0.081i$
                  & $0.200 \pm 0.006$ & $0.098 \pm 0.005$ \\ 
           &   &  & $0.200 \pm 0.002$ & $0.095 \pm 0.001$ \\ 
\tableline
\end{tabular}
\end{center}
\caption{Predicted and estimated eigenfrequencies of $m=1$ modes for disks with 
three different power-law indices $\beta$.}
\label{table:resultsone}
\end{table}

\subsection{Discussion}
\label{sec:discussionone}

The reason for the doublet of $m=1$ modes is that there are two possible paths 
from trailing to leading waves in the inner disk (Toomre 1998, private 
communication). This is because the reflection of an in-going trailing wave off 
the inner cut-out returns some short-leading and long-trailing signal.  Both 
paths complete the feedback loop through swing amplification at corotation.  As 
$Q$ increases, the distinction between the two paths diminishes, and the 
splitting of the doublet decreases.  When $Q \sim 1.5$, linear theory recovers a 
single $m=1$ mode.

\begin{figure}[t]
\centerline{\psfig{file=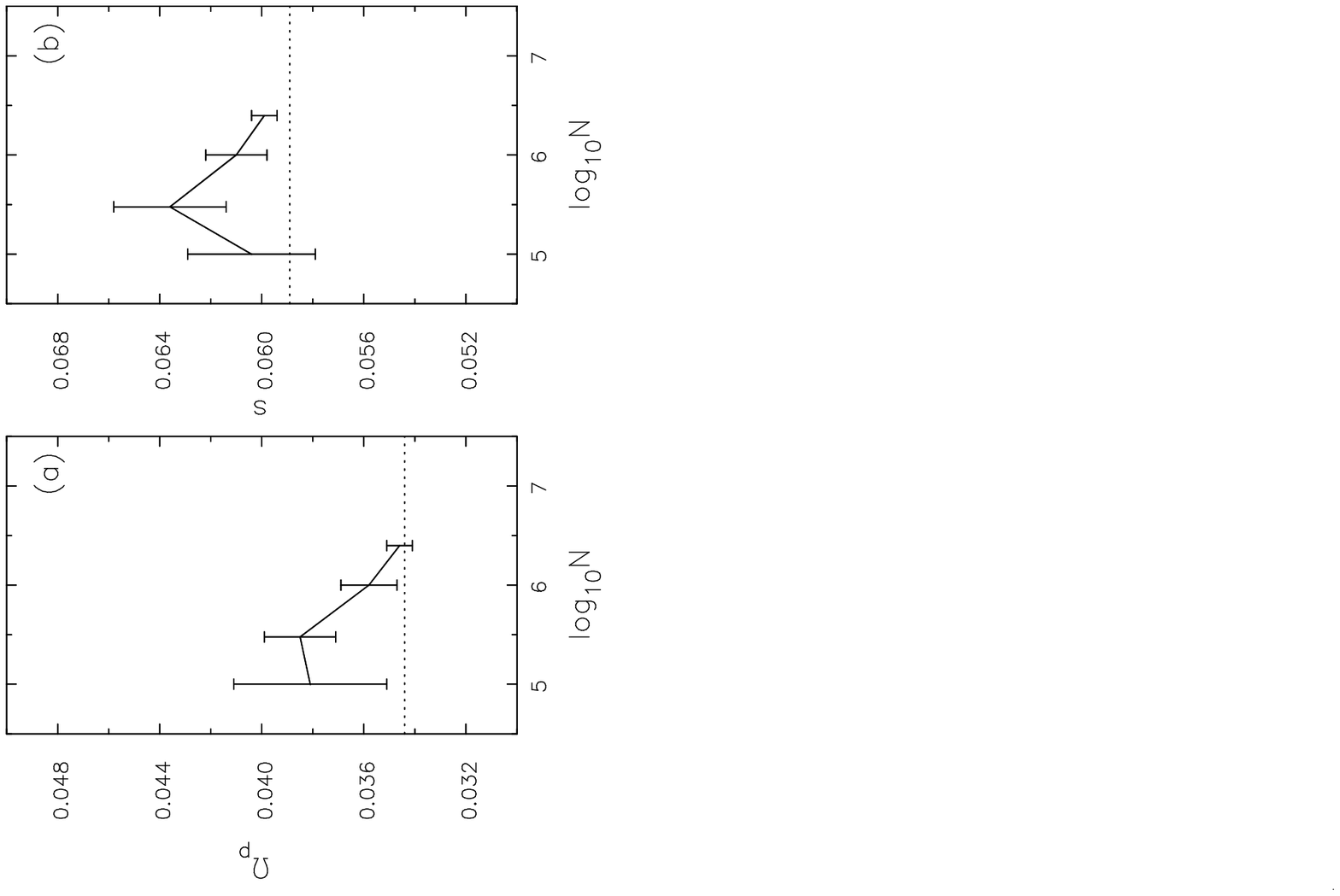,width=.7\hsize,angle=270}}
\caption{The points with error bars show estimates from simulations of (a) the 
pattern speed and (b) the growth rate of the dominant lop-sided mode in a 
$\Qs=1.5$ \Mestel\ disk.  The prediction from linear theory is shown by the 
dotted lines.}
\label{fig:success}
\end{figure}

In order to verify that the discrepancies between predictions and simulations 
are due to the existence of the closely spaced doublet, we have tested the 
prediction of an isolated $m=1$ mode in a $Q_s=1.5$ \Mestel\ disk.  Unlike all 
results reported above, in this case the data from a simulation are well 
approximated by a single coherent mode, with little interference from noise.  
Furthermore, we no longer find significant variations in the measured frequency 
between different draws of particles when $N = 2.5 \times 10^6$.  Figure 
\ref{fig:success} shows that in this clean case we are able to achieve truly 
convincing agreement between linear theory and simulation -- the discrepancy is 
less than 2\% for the largest $N$ we have used.  This result suggests that in 
this case almost all the in-going trailing wave reflects to a leading wave, 
thereby allowing the mode to outgrow any interference from particle noise and to 
stand out quickly and cleanly.  The cleaner reflection at $m=1$ is probably 
largely due to the greater spatial scale of the mode, but the absence of an ILR 
for $m=1$ waves may also be significant, since no residual damping is possible.

\section{A Completely Stable Disk Model}
\label{sec:stable}

\subsection{Linear Stability Theory}

\begin{figure}[t]
\centerline{\psfig{file=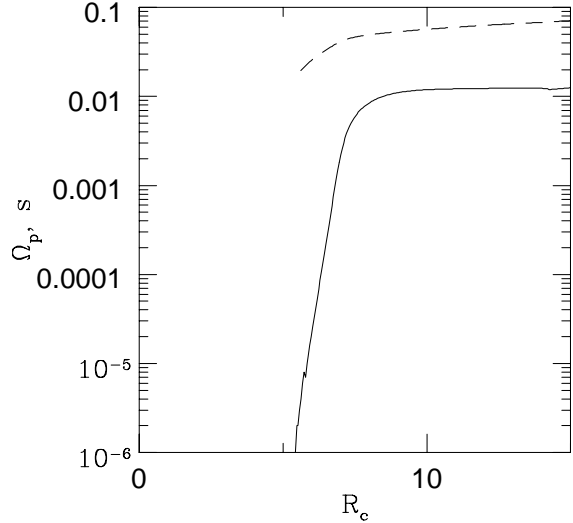,width=0.5\hsize}}
\caption{The pattern speed $\Omegap$ (dashed line) and growth rate $s$ (full 
line) of the dominant $m=1$ mode as the position of the outer taper $\Rc$ is 
varied.  In this $\Qs=1.0$ power-law disk with $\beta = 0.05$ ,$\nu=1$ and 
$\mu=2$, the model is stable against one-armed instabilities when $\Rc \lta 5$.}
\label{fig:ellmodes}
\end{figure}

Our aim here is to provide a clear-cut example of a stable, massive disk with a 
nearly exponential radial profile.  We adopt a power-law disk with a gently 
falling rotation curve ($\beta = 0.05$), but choose the inner cut-out index $\nu 
= 1$ and the outer taper index $\mu=2$.  Figure~\ref{fig:ellmodes} shows how the 
pattern speed and the growth rate of the dominant one-armed mode depends on the 
characteristic radius of the outer taper $\Rc$ for a disk of nominal $\Qs = 
1.0$.  [Note that the nominal $\Qs$ value given in 
equation~(\ref{eq:localsigmin}) refers to the underlying scale-free disk, 
whereas the actual $Q$ value of the tapered disk is higher.]  
Figure~\ref{fig:ellmodes} shows that when the outer taper is centered on $\Rc 
\lta 5$, this disk becomes completely stable to one-armed modes.  We have 
verified that this disk is also linearly stable to the $m=2,\,3$ and $4$ modes.  
At least within the framework of linear theory, the strongly tapered disk is 
completely stable, therefore.

\begin{figure}[p]
\centerline{\psfig{file=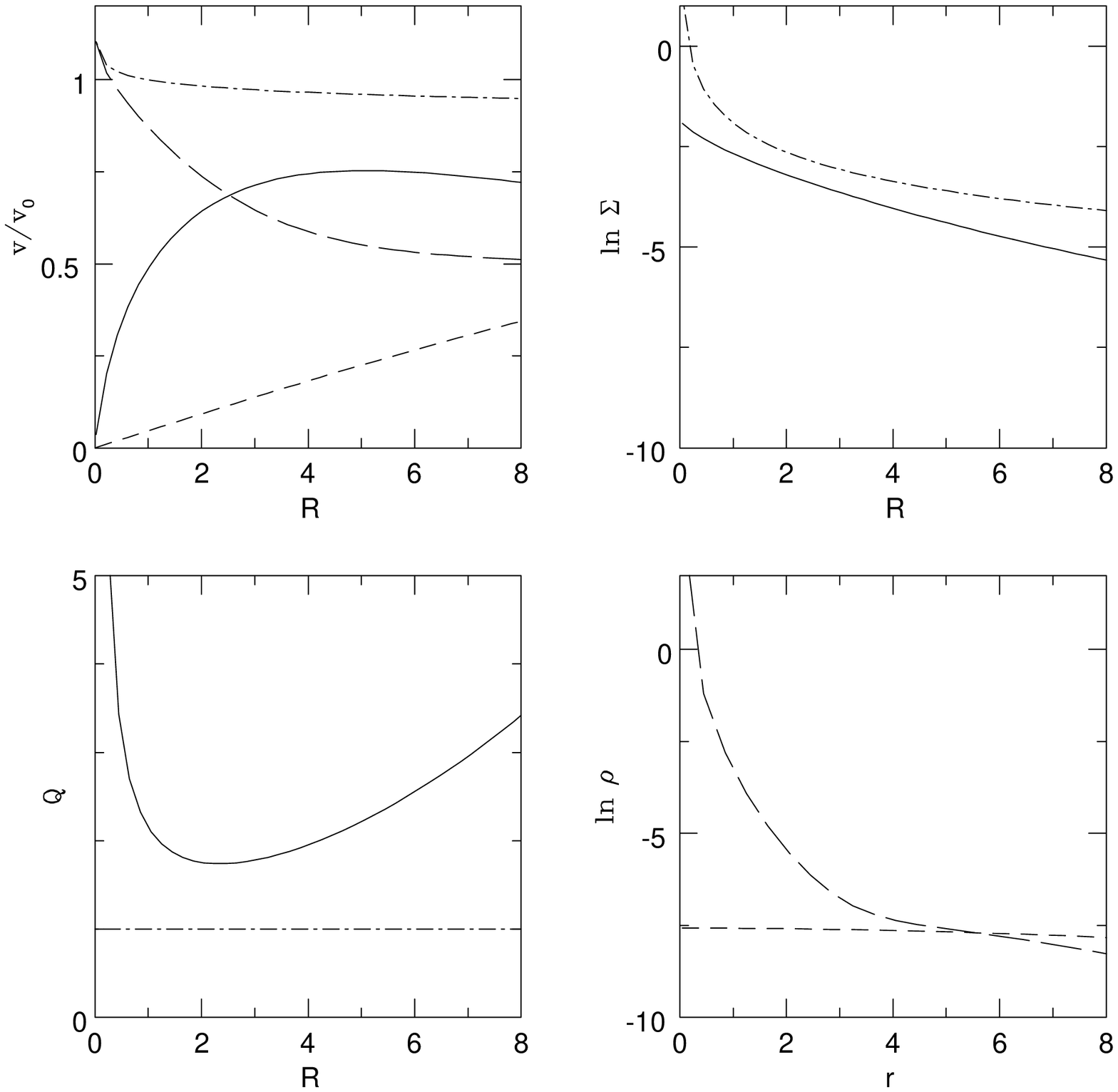,width=0.9\hsize}}
\caption{(a) The rotation curves of the disk (unbroken line), bulge 
(long-dashed) and halo (short-dashed) are plotted against cylindrical radius 
$R$. The total rotation curve (dot-dashed) is gently falling. Throughout the 
inner galaxy, the disk and bulge are the major contributors to the rotation 
curve.  (b) The radial variation of the disk surface density (unbroken line) 
which is quasi-exponential.  The dot-dashed line shows the surface density of 
the \Mestel\ disk for comparison. (c) The true $Q$ profile of the disk (unbroken 
line) is shown against radius $R$.  It rises sharply near the center and in the 
outer parts,although there is an intermediate regime where $Q \sim 1.75$.  The 
dot-dashed line shows the nominal $\Qs = 1.0$. (d) The density of the spherical 
bulge (long-dashed line) and halo (short-dashed line) are shown against 
spherical radius $r$.}
\label{fig:modelpanels}
\end{figure}

\subsection{A quasi-exponential disk}

\label{sec:qexpdisk}
We have removed material from the disk with inner and outer tapers without 
changing the potential.  The extra central attraction is provided by 
unresponsive matter, which we think of as two rigid spherical components, a 
bulge and halo, whose density distributions are determined as follows.

The disk surface density has a logarithmic spiral decomposition:
\begin{equation}
\Sigma (R) = \Sigma_0 \int_{-\infty}^{\infty} d\alpha B(\alpha)
\exp\left[\left (i\alpha-\ffrac{3}{2}\right)\log \left({R\over
R_0}\right) \right],
\end{equation}
where $B(\alpha)$ is given by 
\begin{equation}
B(\alpha) = {1\over 2 \pi} \int_0^\infty {\Sigma (R)\over
\Sigma_0} \exp \left[ -\left(i\alpha -\ffrac{3}{2}\right)\log \left( 
{R\over R_0} \right) \right] {dR \over R}
\label{eq:logtransform}
\end{equation}
For the power-law disks, the integration is analytic; the details are given in 
Appendix A. With this decomposition in hand, we find the spherical density 
$\rhod$ that reproduces the forces in the equatorial plane provided by the disk 
is
\begin{equation}
\rhod (r) = {\Sigma_0\over R_0}\left({r \over R_0}\right)^{-\ffrac{5}{2}}
\int_0^\infty d\alpha (\alpha^2 + \ffrac{1}{4}) K(\alpha,0) B(\alpha)
\exp\left[ i \alpha \log\left( {r\over R_0}\right) \right],
\label{eq:diskdensity}
\end{equation}
where $K(\alpha,0)$ is the axisymmetric Kalnajs (1971) gravity function.  
Therefore, the spherical bulge density $\rhob$ and the halo density $\rhoh$ must 
satisfy
\begin{equation}
\rhob (r) + \rhoh (r) = {v_0^2 (1-\beta)\over 4\pi G r^{2+\beta}} - \rhod (r),
\end{equation}
with $\rhod$ on the right hand side given by equation~(\ref{eq:diskdensity}).  
There is some freedom in the partition of the rigid density between the bulge 
and the halo.  We choose the dark halo to be a cored power-law sphere with the 
potential-density pair (\eg\ Evans 1993)
\begin{equation}
\rhoh (r) = {v_0^2 R_0^\beta (r^2(1-\beta) + 3 \rh^2)\over 4\pi G(r^2 +
\rh^2)^{2+\beta/2}},\qquad\qquad \psih (r) = {v_0^2\over\beta} {R_0^\beta
\over (\rh^2 + r^2)^{\beta/2}}.
\end{equation}
To ensure the halo makes a modest contribution to the rotation curve in the 
inner galaxy, we chose the core radius to be large, $\rh = 20$, giving it an 
almost uniform density in the inner parts.  At large radii, on the other hand, 
the halo provides almost all of the rotational support.  The remainder of the 
cut-out density is ascribed to a bulge.

The equilibrium properties of this model are shown in 
Figure~\ref{fig:modelpanels}.  Panel (a) shows the rotation curves of the disk, 
bulge and halo components.  The whole model has a slowly falling rotation curve. 
 The active disk has a quasi-exponential profile (panel b). The distribution 
functions given by equation~(\ref{eq:fcutouteq}) may be the simplest known which 
yield quasi-exponential disks with flat rotation curves.  As the true $Q$ 
profile of the disk (panel c) is always larger than unity, the disk is 
axisymmetrically stable.  At both small and large radii, $Q$ rises sharply where 
a small fraction of the mass remains active, while $Q\lta2$ at intermediate 
radii.  Panel (d) shows the density of the spherical bulge and halo.  As the 
halo makes a negligible contribution to the mass budget in the inner parts, we 
can regard this is a ``maximum disk'' model, albeit with a massive bulge.

It should be noted that although the surface density 
(Figure~\ref{fig:modelpanels}b) reaches at most barely more than half that of 
the untapered disk near $R \sim 2$, the tapered disk's contribution to the 
circular speed (Figure~\ref{fig:modelpanels}a) peaks at some 75\% of that from 
the full-mass disk near $R \sim 5$.  This result should not be surprising since 
the taper has cut away material at larger radii which previously depressed the 
disk's contribution to the central attraction.

\begin{figure}[p]
\centerline{\psfig{file=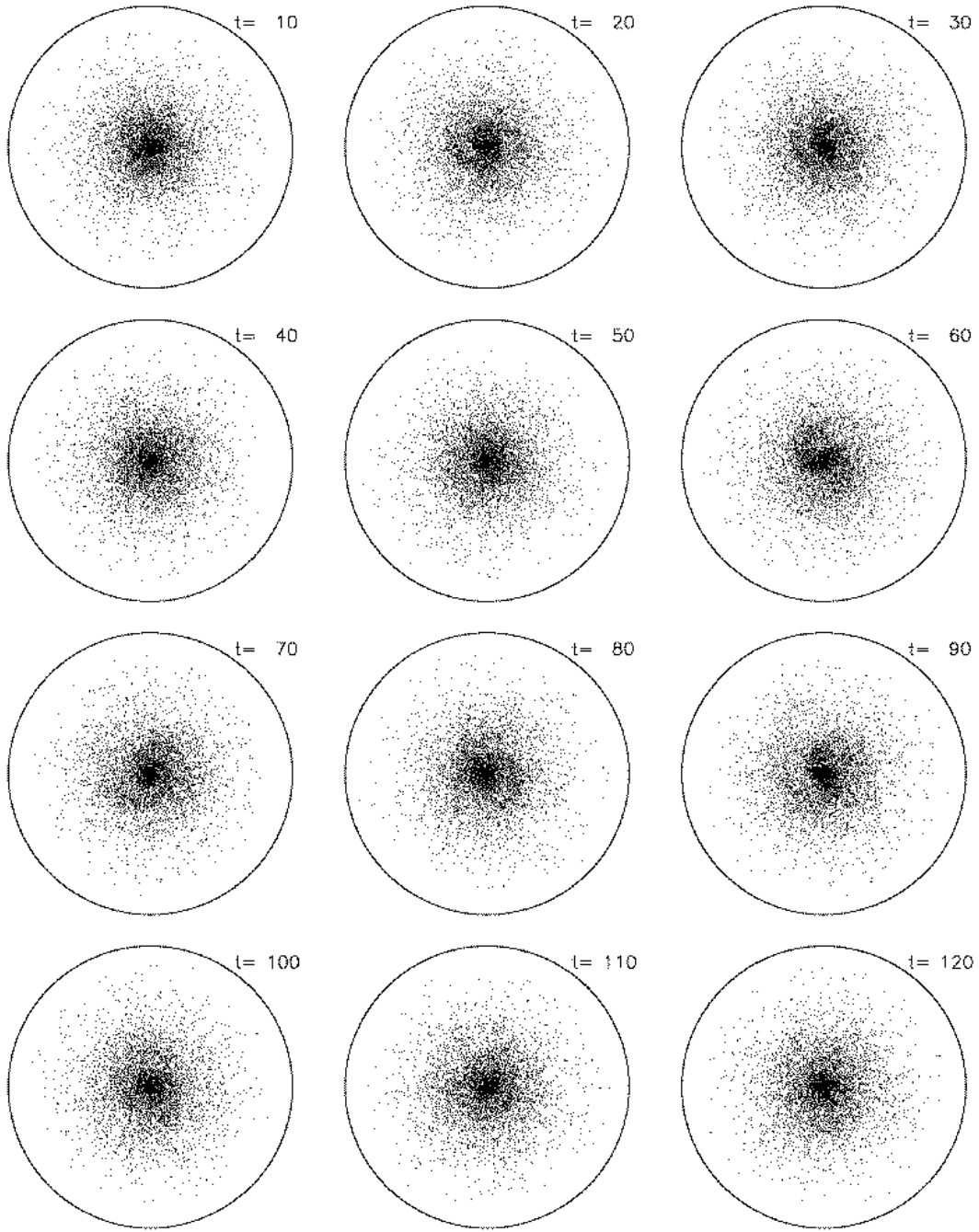,width=.9\hsize}}
\caption{The positions of one particle in 10 from a noisy start simulation of 
model described in \S\ref{sec:qexpdisk}.  The radius of the circles is $17.2R_0$ 
and the times are in units of $R_0/v_0$.  The simulation employed $N=5 \times 
10^4$ and forces from $m=0 - 4$ were all active.  The model appears to be stable 
to bar-forming and lop-sided modes, but as usual swing-amplified noise quickly 
produces spiral patterns which heat the disk, especially in the inner parts.}
\label{fig:stable}
\end{figure}

\subsection{Simulations}
We have used our $N$-body code to test the linear theory prediction that this 
model has no global instabilities.  Figure \ref{fig:stable} shows the result 
from a noisy start simulation with a modest number of particles ($N=5 \times 
10^4$) and with $m=0-4$ forces calculated from the particles.  All other 
numerical parameters were as given in Table \ref{table:jerrysparameters}.  The 
force acting on each particle at every step included the fixed central 
attraction of our bulge/halo in addition to the forces computed from the 
particles.

The simulation quickly develops some spiral patterns, which result from 
swing-amplified particle noise, but there is no evidence for a coherent inner 
bar or any lop-sidedness.  A test with ten times more particles behaved 
similarly, but the spirals were so weak that the disk barely heated at all, and 
no long-lived coherent waves were detectable at any $m$.

We conclude that this model is robustly stable.

\section{Conclusions}

It should no longer be doubted that real galaxies can be stabilized by dense 
centers, as argued by Toomre (1981).  We have finally confirmed Zang's (1976) 
global stability analysis for the \Mestel\ disk, and also its generalization to 
other power laws by Evans \& Read (1998b).  This encouraging agreement between 
linear theory and simulations for these idealized galaxy models buttresses 
Sellwood's (1985, 1999) claims of global stability for more realistic, but less 
analytically tractable, models with hard centers.

The lop-sided ($m=1$) instabilities of the power-laws disks, first discovered by 
Zang, can be controlled by reducing the disk surface density to $\lta 2/3$ of 
that of a full-mass disk, without changing the potential (Toomre 1981), much in 
the same manner as bars can be avoided in soft-centered disks by making them 
sub-maximal.  In this paper, we have been able to show that the $m=1$ mode can 
also be avoided by tapering the massive outer disk until its surface density 
declines in a quasi-exponential manner.  The mechanism is no different since the 
surface density reaches barely more than half that of the untapered disk at any 
radius, but the result is a more realistic model in which the disk contribution 
to the central attraction in the inner parts is decreased to a smaller extent 
than is the mass.

Many real galaxies do appear to be stabilized by dense centers.  High-resolution 
kinematic observations of spiral galaxies by both Rubin, Kenney \& Young (1997) 
and by Sofue \etal\ (1999) have revealed high orbital speeds close to their 
centers.  In the cases reported by Sofue \etal, for example, the only galaxies 
with gently rising rotation curves all have rather low luminosity (a peak 
circular velocity $< 150 \hbox{ km s}^{-1}$).  Such galaxies are required to 
have large dark matter fractions (\eg\ Broeils 1992) -- presumably enough to 
suppress bar-forming instabilities.  Almost every galaxy with a circular speed 
in excess of 150 km s$^{-1}$ has a steep inner rise in the rotation curve, and 
must therefore be bar-stable whatever its dark matter content.

The linear modes of these power-law disks proved to be much harder to identify 
than those in previous studies.  The predicted bisymmetric instabilities are 
completely obscured by swing-amplified noise unless we employ much larger 
numbers of particles than are needed in soft-centered disks, and even then the 
underlying mode does not stand out clearly.  We attribute this difference to the 
different reflection mechanism in the feed-back loop for the power-law disks.  
The mode stands out clearly from the noise only in cases where the inner 
reflection is total, as in soft-centered disks, and possibly also for $m=1$ 
modes in hard-centered disks (\S5.2).  But particle noise readily swamps the 
weakened modes which result when the reflection involves considerable leakage, 
as happens for the bisymmetric modes in hard-centered disks (\S4.2).

We have been able to confirm the predicted most unstable $m=1$ mode in the 
$\Qs=1.5$ disk with a precision approaching that obtained by Earn \& Sellwood 
(1995) for the bar modes in the isochrone disk.  Our simulations of the cooler 
$\Qs=1$ disk appeared consistent with only a single mode where linear theory 
predicted a closely-spaced doublet; nevertheless, the frequency we obtained was 
always some ``average'' of the predicted pair.

We have shown that one-armed instabilities are tamed by tapering the outer disk. 
 We note that the lop-sided instability has implications for the dark matter 
candidate proposed by Pfenniger, Combes \& Martinet (1994).  They suggested that 
flat rotation curves result from a high surface density of very cold molecular 
gas in a rotationally supported disk extending beyond the optical edge.  This 
hypothesis is ruled out on stability grounds: massive extended disks with flat 
rotation curves are grossly unstable to $m=1$ modes.

We have presented a realistic galaxy model that is globally stable.  It has a 
quasi-exponential disk in an almost flat rotation curve potential, and includes 
both a halo with a large core radius and a bulge.  The model is stabilized 
against bar modes by its dense center and against one-armed modes by the 
exponentially decreasing surface density profile.  Even though the halo makes an 
negligible contribution to the central attraction in the inner parts, the model 
is clearly globally stable as we demonstrate using both normal mode analysis and 
$N$-body simulations.

\acknowledgments We are indebted to Alar Toomre for his comments on an early 
draft of this paper and detailed, patient explanations of the mode mechanisms.  
Useful conversations with David Earn are also acknowledged.  We thank the 
referee (Agris Kalnajs) for a thoughtful and stimulating report. This work was 
begun at the Isaac Newton Institute, Cambridge and continued at the Lorentz 
Centre, Leiden.  The hospitality of both places is acknowledged.  JAS is 
supported by NSF grant AST 96/17088 and NASA LTSA grant NAG 5-6037.  NWE thanks 
the Royal Society for financial support.

\appendix
\section{The Density Transform}

In section~\ref{sec:qexpdisk}, we exploit the decomposition into logarithmic 
spirals to find the three-dimensional spherical bulge density that provides the 
forces to compensate for the matter carved out of our disks. The integration for 
the logarithmic spiral density transform~(\ref{eq:logtransform}) can be carried 
through explicitly for the power-law disks. This is done by writing the density 
as an integral over velocity space of the cut-out distribution function, and 
then reversing the orders of integrations.

For the case of our quasi-exponential disk, the inner cut-out is $\nu=1$ while 
the outer taper is $\mu=2$. The transform then is
\begin{eqnarray}
B(\alpha) & = & {1\over 2 (2 + \beta)}\left( {2\over \beta}
\right)^\betah {B(2 + 1/\beta +\gamma/\beta, \betah) \over B(1/2 +
\gamma/2, \betah)} {1 \over 1 + \eta^{2(2+\beta)/(2-\beta)}} \nonumber\\ 
& & \times  \left\{ \eta^{2\betah + (2+\beta)/(2-\beta)} \sec\,
\alphah +\eta^{2\betah} \cosec\, \alphah - 2\cosec\, 2\alphah \right\}, 
\end{eqnarray}
where $\eta = r_0v_0 /\Lc$ and
\begin{equation}
\alphah = {\pi (\shalf - \beta - i \alpha) \over 2 + \beta},\qquad
\betah = {\beta + i \alpha - \shalf \over 2 - \beta}.
\end{equation}
This is valid for $\beta >0$; slightly different expressions hold good for 
$\beta \le 0$.

\end{document}